\documentclass[prb,aps]{revtex4}

\newcommand{\ba}{\begin{eqnarray}}
\newcommand{\ea}{\end{eqnarray}}

\usepackage{amsmath}

\begin{document}

\title{Inaccurate use of asymptotic formulas}

\author{Rados\l aw Maj}
\email{radcypmaj@poczta.onet.pl}

\affiliation{Institute of Physics, \'Swi\c etokrzyska Academy,
ul. \'Swi\c etokrzyska 15, PL - 25-406 Kielce, Poland}

\author{Stanis\l aw Mr\' owczy\' nski}
\email{mrow@fuw.edu.pl}

\affiliation{So\l tan Institute for Nuclear Studies,
ul. Ho\.za 69, PL - 00-681 Warsaw, Poland \\
and Institute of Physics, \'Swi\c etokrzyska Academy,
ul. \'Swi\c etokrzyska 15, PL - 25-406 Kielce, Poland}

\begin{abstract}
The asymptotic form of the plane wave decomposition into spherical waves,
which is used to express the scattering amplitude in terms of  phase
shifts, is incorrect. We explain why and show how to circumvent the
mathematical inconsistency.
\end{abstract}

\maketitle


In quantum mechanics the following plane wave decomposition into
spherical waves is used
\begin{equation}
\label{decomp1}
e^{i{\bf k} \cdot {\bf r}} = \sum_{l=0}^{\infty} i^l (2l +1)
P_l(\cos \Theta)
 j_l(kr),
\end{equation}
where ${\bf k} \cdot {\bf r} = kr \cos\Theta$, $ P_l$ is the
$l$th Legendre polynomial, and $j_l$ is the $l$th spherical Bessel function.
Because we are interested in the large distance behavior of the wave
function in scattering theory, we need the asymptotic form of the spherical 
function 
\begin{equation}
 \label{asymp1} j_l(kr) \approx \frac{\sin(kr -\pi l/2)}{kr},
\end{equation}
and we rewrite Eq.~(\ref{decomp1}) as
\begin{equation}
\label{decomp2} e^{i{\bf k} \cdot {\bf r}} 
\buildrel ? \over \approx \frac{1}{kr}\sum_{l=0}^{\infty} i^l (2l +1)\,
P_l(\cos\Theta) \sin(kr -\pi l/2) .
\end{equation}
Equation~(\ref{decomp2}) is given in numerous textbooks on quantum 
mechanics, including Schiff\cite{Schiff} and
Landau and Lifshitz.\cite{LL} Astonishingly, the expression (\ref{decomp2}) 
is meaningless, and for this reason we put the question mark over the 
approximate equality. The series is not only divergent, but it cannot
even be treated as an asymptotic expansion of the function 
$e^{i{\bf k} \cdot {\bf r}}$ at large distances.

To see how badly the series (\ref{decomp2}) diverges, we consider 
the special case $\cos\Theta =1$. Then, $P_l(1)=1$, and after 
a simple calculation we obtain
\begin{subequations}
\label{diver}
\ba
e^{ikr} 
& \buildrel ? \over \approx &
\frac{\sin(kr)}{kr} 
\sum_{l=0}^{\infty} (2l +1) \cos ^2(\pi l/2)
-i \frac{\cos(kr)}{kr} 
\sum_{l=0}^{\infty} (2l +1) \sin^2(\pi l/2)
\\ 
&=& \frac{\sin(kr)}{kr} 
\sum_{n=0}^{\infty} (4n + 1) 
- i \frac{\cos (kr)}{kr} 
\sum_{n=0}^{\infty} (4n + 3) .
\ea
\end{subequations}
According to Eq.~(\ref{diver}), both the real and imaginary parts
of $e^{ikr}$ contain a divergent series, for any value of $r$.

However, we often consider asymptotic 
series that are divergent, but still correctly represent certain 
functions. The infinite series 
$a_0(x) + a_1(x) + a_2(x) + \dots$ is the asymptotic expansion of 
the function $f(x)$ at $x_0$ (which can be infinite) if\cite{Copson}
\begin{equation}
 \label{def-asym}
\frac{1}{a_n(x)} \Big(f(x) - \sum_{l=0}^n a_l(x)\Big) \rightarrow 0 
\quad \mbox{for}\ x \rightarrow x_0 .
\end{equation}
Equivalently, the series is asymptotic if
\begin{equation}
 \label{cond-asym}
\frac{a_{l+1}(x)}{a_l(x)} \rightarrow 0 \quad \mbox{for}\
x \rightarrow x_0 .
\end{equation}
Due to the definition (\ref{def-asym}), any finite subseries 
of an asymptotic series approximates the function $f(x)$ and 
the approximation becomes better and better as 
$x \rightarrow x_0$. However, the series (\ref{decomp2}) does 
not satisfy the condition (\ref{cond-asym}), and consequently, 
it cannot be treated as an asymptotic expansion of $e^{i{\bf k} \cdot
{\bf r}}$ at large distances.

What is wrong with the expansion (\ref{decomp2})? It appears that
the approximate formula (\ref{asymp1}) requires that
\begin{equation}
 \label{condition}
kr \gg \frac{1}{2} l(l+1) .
\end{equation}
For completeness, we derive this condition here, and find not only the 
first but also the second term of the $1/z$ expansion of $j_l(z)$. It 
is well known (see, for example, Ref.~\onlinecite{Korn}) that the spherical
Bessel functions can be written as
\begin{equation}
\label{bessel}
j_l(z) = z^l \Big(- \frac{1}{z} \frac{d}{dz} \Big)^l
\frac{\sin z}{z} .
\end{equation}
If we use Eq.~(\ref{bessel}) and the recursion formula,
\begin{equation}
j_{l+1}(z) = - z^l \frac{d}{dz}
\Big(\frac{1}{z^l} j_l(z) \Big) ,
\end{equation}
we can easily prove by induction that
\begin{equation}
 \label{asymp2}
j_l(z) = \frac{\sin(z -\pi l/2)}{z}
+ \frac{1}{2} l(l+1) \frac{\cos(z -\pi l/2)}{z^2}
+ {\cal O}(\frac{1}{z^3}) .
\end{equation}
If we compare the two terms of the expansion (\ref{asymp2}), we find
that the approximation (\ref{asymp1}) holds if the condition
(\ref{condition}) is satisfied. When we perform the summation 
in Eq.~(\ref{decomp2}), we find that the terms for sufficiently large $l$
violate the requirement (\ref{condition}), and effectively destroy even 
the approximate equality.

Although the decomposition (\ref{decomp2}) is incorrect, the results 
obtained by means of it are usually correct. Obviously, the famous 
formula, that expresses the scattering amplitude
via the phase  shifts, is correct. However, it is of interest to see why the
derivation works. Therefore, we first discuss the standard procedure, which
can be  found, for example, in Refs.~\onlinecite{Schiff} and \onlinecite{LL},
and then we show how to avoid the mathematical inconsistency.

By assuming azimuthal symmetry, the scattered wave 
function is
\begin{equation}
 \label{wave1}
\phi_{\bf k}({\bf r}) = \sum_{l=0}^{\infty} A_l i^l (2l +1) P_l(\cos\Theta) 
R_l(r) ,
\end{equation}
where the $R_l$ are the radial wave functions and the $A_l$ are
coefficients to be determined.
If we assume that the asymptotics of the radial functions are 
\begin{equation}
 \label{asymp-rad}
R_l(r) \approx \frac{\sin(kr -\pi l/2+ \delta_l)}{kr} ,
\end{equation}
where $\delta_l$ denotes the $l$th phase-shift, we can rewrite
Eq.~(\ref{wave1}) as 
\begin{equation}
 \label{wave2}
\phi_{\bf k}({\bf r}) 
\buildrel ? \over \approx
\frac{1}{kr}\sum_{l=0}^{\infty} A_l i^l (2l +1) P_l(\cos \Theta) \sin(kr
-\pi l/2 + \delta_l) .
\end{equation}
We still put the question mark over the equalities that are
mathematically inappropriate. 

Now, we compare the wave function (\ref{wave2}) with 
the expected asymptotic form of the scattered wave function
\begin{equation}
 \label{scatt}
\phi_{\bf k}({\bf r}) = e^{i{\bf kr}} + f(\Theta ) 
\frac{e^{ikr}}{r} ,
\end{equation}
where $f(\Theta )$ is the scattering amplitude. If we use the 
plane-wave decomposition (\ref{decomp2}), we find the 
equation
\ba
\label{compar}
&&\frac{1}{kr}\sum_{l=0}^{\infty} A_l i^l (2l +1) P_l(\cos \Theta) \sin(kr
-\pi l/2 + \delta_l) \nonumber \\
{}&& \buildrel ? \over 
\approx
\frac{1}{kr}\sum_{l=0}^{\infty} i^l (2l +1) P_l(\cos\Theta) \sin(kr -\pi l/2)
+ f(\Theta ) \frac{e^{ikr}}{r} .
\ea
If we equate the terms proportional to $e^{-ikr}$, we find that
$A_l = e^{i \delta_l}$, which when substituted into the terms 
proportional to $e^{ikr}$, provides the well known result
\begin{equation}
 \label{ampli}
f(\Theta ) = 
\frac{1}{2ik}\sum_{l=0}^{\infty}(2l +1) P_l(\cos\Theta)
\big[e^{2i \delta_l} - 1 \big] .
\end{equation}
We obtained the relation (\ref{ampli}) using the 
mathematically meaningless equations 
(\ref{decomp2}), (\ref{wave2}), and (\ref{compar}). Next, we show how to
derive Eq.~(\ref{ampli}) avoiding the inconsistency.

Again we start with the wave function in the form 
(\ref{wave1}), and we use the asymptotics of the radial wave function
(\ref{asymp-rad}), but only for fixed values of $l$. For this
reason we calculate the projection
\begin{equation}
 \label{project1}
\int_{-1}^{+1}d (\cos\Theta) \phi_{\bf k}({\bf r}) P_l(\cos\Theta)
= 2 i^l A_l R_l(r) ,
\end{equation}
where we have taken into account that the Legendre polynomials are orthogonal
\begin{equation}
\int_{-1}^{+1}d (\cos\Theta ) 
 P_l(\cos\Theta)
 P_{l'}(\cos\Theta)
= \frac{2}{2l + 1} \delta^{ll'} .
\end{equation}
We decompose the scattering amplitude,
\begin{equation}
 \label{ampli-decomp}
f(\Theta ) = \frac{1}{2ik}\sum_{l=0}^{\infty} C_l i^l 
(2l +1) P_l(\cos\Theta),
\end{equation}
and we project the scattered wave function (\ref{scatt}) as
\begin{equation}
 \label{project2}
\int_{-1}^{+1}d (\cos\Theta ) 
\Big(e^{i{\bf kr}} + f(\Theta ) \frac{e^{ikr}}{r} \Big)
 P_l(\cos\Theta)
= 2 \, i^l \Big(j_l(kr) + C_l \frac{e^{ikr}}{r} \Big) .
\end{equation}
Next, we equate the asymptotic forms of the projections 
(\ref{project1}) and (\ref{project2}), and thus, instead of
Eq.~(\ref{compar}), we obtain 
\begin{equation}
A_l \sin(kr -\pi l/2 + \delta_l)
= \sin(kr -\pi l/2) + C_l k \frac{e^{ikr}}{r} .
\end{equation}
We compare the terms proportional to $e^{-ikr}$ and $e^{ikr}$,
respectively, and find that $A_l = e^{i \delta_l}$ and
\begin{equation}
C_l = \frac{1}{2ik} e^{-i\pi l/2} \big[e^{2i \delta_l} - 1 \big] ,
\end{equation}
which, due to Eq.~(\ref{ampli-decomp}), again provides the
correct result (\ref{ampli}).

Although the problem discussed here looks pure academic it was discovered 
in the course of concrete calculations. To
simplify the calculation of a correlation function where the scattering wave
function enters, we used the form (\ref{wave2}) with $A_l = e^{i \delta_l}$
as is given in many books. We were interested in the complete sum of partial
waves, and we used Eq.~(\ref{wave2}) instead of Eq.~(\ref{scatt}) to exploit
the orthogonality of Legendre polynomials. Needless to say the calculation
went wrong, showing that the asymptotic expressions must be treated very
carefully.

\acknowledgements
We are very grateful to Konrad Bajer and Iwo Bia\l ynicki-Birula
for correspondence and stimulating criticism.

\end{document}